\documentclass[aps,prb,twocolumn,floatfix,showpacs,superscriptaddress]{revtex4}

\usepackage{graphicx}
\usepackage{bbm}
\usepackage{amsmath,amssymb}

\newcommand{\be}{\begin{equation}}
\newcommand{\beq}{\begin{equation}}
\newcommand{\ee}{\end{equation}}
\newcommand{\bea}{\begin{eqnarray}}
\newcommand{\eea}{\end{eqnarray}}
\newcommand{\ba}{\begin{array}}
\newcommand{\ea}{\end{array}}

\renewcommand{\vr} {{\bf r}}

\begin{document}
\title{Simple exchange-correlation potential with 
a proper long-range behavior for low-dimensional nanostructures}
\author{S. Pittalis}
\email[Electronic address:\;]{pittaliss@missouri.edu}
\affiliation{Department of Physics and Astronomy, University of Missouri, Columbia, Missouri 65211, USA}
\author{E. R{\"a}s{\"a}nen}
\email[Electronic address:\;]{erasanen@jyu.fi}
\affiliation{Nanoscience Center, Department of Physics, University of
  Jyv\"askyl\"a, FI-40014 Jyv\"askyl\"a, Finland}

\date{\today}

\begin{abstract}
The exchange-correlation potentials stemming from the 
local-density approximation and several generalized-gradient 
approximations are known to have incorrect asymptotic
decay. This failure is independent of the dimensionality, but so far the
problem has been corrected -- within the mentioned approximations -- 
only in three dimensions.
Here we provide a cured exchange-correlation potential in
two dimensions, where the applications have a continuously 
increasing range in, e.g.,
semiconductor physics. The given potential is a 
generalized-gradient approximation,
which is as easy to apply as the local-density approximation.
We demonstrate that the corrected potential agrees very well with the
analytic result of a two-electron quantum dot in the asymptotic
regime, and yields plausible exchange-correlation potentials for
larger two-dimensional systems.
\end{abstract}

 
\maketitle

\section{Introduction}

Developments in density-functional theory~\cite{dft} (DFT), particularly
in the derivation of functionals for exchange-correlation (xc)
energies and potentials,~\cite{xc_review} can often be
characterized by a compromise between accuracy and efficiency.
In quantum chemistry and condensed matter physics, applications of 
generalized-gradient approximations (GGAs) are still among the most 
popular choices to evaluate the xc terms in the Kohn-Sham (KS) scheme.
Even the local-density approximation (LDA) has well preserved 
its popularity due to the several beneficial properties, e.g.,
the computational efficiency, the natural compatibility between the
exchange and correlation, satisfaction of the sum rule of the xc 
hole etc.~\cite{dft}

One of the most prominent problems in the LDA and in many  
GGAs is the wrong asymptotic behavior of the xc potential. 
Essentially, this is due to the fact that these approximations 
depend in a oversimplified fashion on the (exponentially decaying)
electron density, leading to the wrong exponential decay of
the approximate xc potential.
A simple and efficient cure for this problem was put 
forward by van Leeuwen and Baerends.~\cite{LB94}  
They provided a GGA for the xc potential with the proper asymptotic
behavior, commonly known as the LB94 functional (or potential).
The LB94 potential is very accurate for, e.g., ionization energies and electron
affinities, and it has received considerable popularity in quantum
chemistry.  

The three-dimensional (3D) LDA and GGAs are not applicable for 
systems having suppressed dimensionality~\cite{kim,pollack,constantin} 
such as the well-known two-dimensional (2D) examples, e.g.,
semiconductor quantum dots (QDs), rings, slabs, and 
quantum Hall systems. Instead,
the 2D LDA~\cite{rajagopal,attaccalite} and
-- to the best of our knowledge -- the {\em only} 2D GGA~\cite{2DGGA}
are suitable, in principle, to deal with those systems. 
However, these functionals suffer
from the same deficiencies as their 3D counterparts, most importantly,
the wrong asymptotic behavior. Motivated by this fact, we derive
here a xc potential which corrects the asymptotic decay and
remains computationally efficient due to its simple GGA form.
In the derivation we essentially follow the steps of 
Ref.~\onlinecite{LB94}. Finally we demonstrate the usefulness of
the model potential by considering a few typical 2D applications:
two semiconductor QDs and two quantum rings.

\section{Corrected long-range behavior}

Following Ref.~\onlinecite{LB94}
we start by expressing the total xc potential in the KS scheme as
\be\label{vxc}
v_{xc} = v^{\rm LDA}_{xc} + \rho^{1/2} F(x)
\ee
where 
\be\label{x1}
x =  \frac{| \nabla \rho |}{ \rho^{3/2}}\;.
\ee
In Eq.~(\ref{vxc}), $F(x)$ together with $x$ can be considered as a 
quantity evaluating the relevance of the density gradient as well as 
its asymptotic behavior. We remind that $x$ has been introduced already 
for the 2D GGA in Ref.~\onlinecite{2DGGA}.
We observe that the term $ \rho^{1/2} F(x)$ scales 
linearly under homogeneously scaling.
As it is well known, the linear scaling applies to 
$v_x$ exactly, but this may not be the case for 
$v_c$. Nevertheless, in the spirit of LB94 we treat $v_c$ 
together with $v_x$. This may be justified considering that
for $r \rightarrow \infty$ the leading contribution to the 
asymptotics of $v_{xc}=v_x+v_c$ is due to $v_x$; in other words,
$v_c$ decays faster.~\cite{LB94,filippi} 
It should be noted, however, that the the present correction 
has effect also elsewhere, i.e., not only in the asymptotic limit.

For the function $F(x)$ in Eq.~(\ref{vxc}) we require 
$F(0) = 0$, so that the total $v_{xc}$ will be given by the LDA  
whenever $\nabla \rho \equiv 0$. In this way, the result for the 
homogeneous electron gas is {\em exactly} recovered. 
This property is useful also when considering inhomogeneous systems,
in view of the success of the LDA.

To the leading order, the model should fulfill~\cite{x1,ring}
\be
v_{xc} \sim - \frac{1}{r}\;,
\label{limit}
\ee 
in contrast to the exponential decay of the LDA potential.~\cite{note1} 
This requirement
translates into the following condition
\be\label{FF}
F(x) \sim   - \frac{1}{r}\rho^{-1/2} \;.
\ee
We remind that far away from the central region of QDs~\cite{note2}
\be\label{rho}
\rho  \sim  e^{-\alpha r^2},~~~~ \alpha > 0\;
\ee
and thus, using Eq.~(\ref{rho}) in Eq.~(\ref{x1}) we find
\be\label{x}
x \sim 2 \alpha r \rho^{-1/2}\;.
\ee
Equations~(\ref{FF}) and (\ref{x}) lead to
\be\label{F0}
F(x) \sim  -\frac{x}{2 \alpha r^2}\;.
\ee
Moreover, since
\be
\ln x \sim \frac{\alpha r^2}{2}\;,
\ee
we may re-write Eq.~(\ref{F0}) as follows:
\be\label{F1}
F(x) \sim - \frac{1}{4} \frac{x}{\ln x}\;.
\ee 
The desired behavior is given by a function
\be\label{F}
F(x) = - \beta \frac{x^2}{1 + 4 \beta x \sinh^{-1}\left(x\right)}\;,
\ee
where
\be
\sinh^{-1}\left(x\right) =  \ln( x + \sqrt{1+x^2} ) \;.
\ee
The parameter $\beta$ can be determined, for example, by 
fitting the model potential against the {\em exact} xc potential 
in a solvable test system. We consider the effects of $\beta$ below.

Apart from the factor of four in the denominator of Eq.~(\ref{F1}),
and the 2D definition of $x$, the resulting model is very similar
to the 3D case.~\cite{LB94} Moreover, as for the 3D case, 
the exact constraints for the exchange-correlation
potential resulting from the translational and rotational invariance
are satisfied.

Finally we point out that it is straightforward to obtain a 
{\em spin-dependent} model potential by simply adding
a spin index to all the quantities but $\beta$. 

\section{Further analysis and applications}

Next we test our model for the xc potential in 
a 2D two-electron QD defined by a radial external potential 
$v_{\rm ext}(r)=\omega^2 r^2/2$ with $\omega=1$ (see Ref.~\onlinecite{units} 
for information about the units).
The ground-state (singlet) density
has an exact analytic expression,~\cite{taut,wensauerthesis}
\begin{eqnarray}
\rho(r) & = & \frac{4}{\pi(\sqrt{2\pi}+3)}\Big\{e^{r^2}(1+r^2/2)+
\frac{1}{2}\sqrt{\pi}e^{r^2/2} \times \nonumber \\ 
& \times & \Big[I_0(r^2/2)+r^2 I_0(r^2/2)+r^2 I_1(r^2/2)\Big]\Big\},
\end{eqnarray}
where $I_0(x)$ is the zeroth-order modified Bessel function of the 
first kind. The exact
KS orbital is given simply by $\varphi(r)=\sqrt{\rho(r)/2}$, and the KS equation
can be inverted to obtain the exact KS potential from
\be
v_{\rm KS}(r) = \epsilon_{\rm KS} + \frac{\nabla^2\varphi(r)}{2\varphi(r)},
\ee
where $\epsilon_{\rm KS}=2$ is the KS eigenvalue.~\cite{taut,filippi}
The exact xc potential is then given by
\be
v^{\rm exact}_{xc}(r) = v_{\rm KS}(r)-v_{\rm ext}(r)-v_{H}(r),
\ee
where $v_H$ is the Hartree potential, which is in fact the
only term having {\em no} analytic expression for the given
analytic density. However, it can be solved numerically with
a good precision. 

\begin{figure}
\includegraphics[width=0.90\columnwidth]{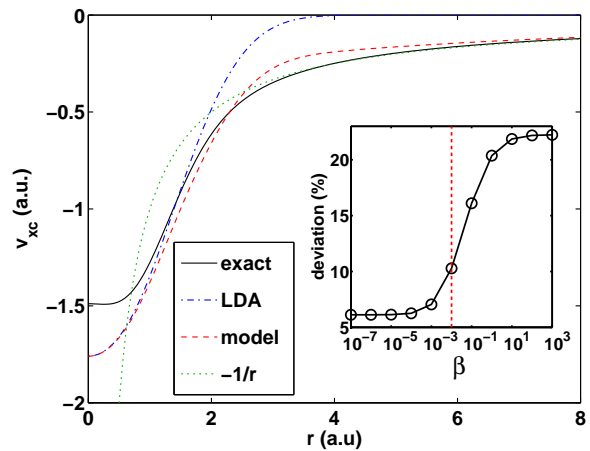}
\caption{(Color online) Exchange-correlation potentials
of a parabolic ($\omega = 1$) two-electron quantum dot calculated
from the analytic density. The result from the 
present model (solid line) has been calculated with
$\beta=0.01$ in Eq.~(\ref{F}). 
The chosen value is kept fixed in the following examples. 
The inset shows the
relative deviation of the integrated model 
potential from the exact result as 
a function of $\beta$ (note the logarithmic scale).
}
\label{fig1}
\end{figure}

In Fig.~\ref{fig1} 
we compare the exact xc potential with the LDA potential 
and with our model, i.e., Eqs.~(\ref{vxc}) and (\ref{F}) with $\beta=0.01$. 
Whereas the LDA potential decays exponentially, the
model potential has the correct asymptotic behavior 
in an excellent agreement with the exact result. At $r=0$
the density gradient is zero and hence the model
agrees with the LDA as required.

To analyze the effect of the parameter 
$\beta$ in Eq.~(\ref{F}) on the xc potential, 
we consider the relative difference
in the integrated xc potential weighted by the density, i.e., 
\be
\Delta I =  \frac{\int d^2 r~  \rho(\vr) \left|v^{\rm exact}_{xc}(\vr)-v_{xc}(\vr)\right|}{\int d^2  r~\rho(\vr) v^{\rm exact}_{xc}(\vr)}\;.
\ee
This quantity is plotted as a function of $\beta$ in the inset
of Fig.~\ref{fig1}. We find that the deviation is
overall very insensitive to $\beta$ (note the logarithmic 
x-axis). We have chosen the value $\beta=0.01$ (dashed line) 
that yields a ``reasonable'' shape for the xc potential 
and a relatively small deviation in $\Delta I$ for this 
test case. The chosen value is kept fixed in the following examples.

\begin{figure}
\includegraphics[width=0.75\columnwidth]{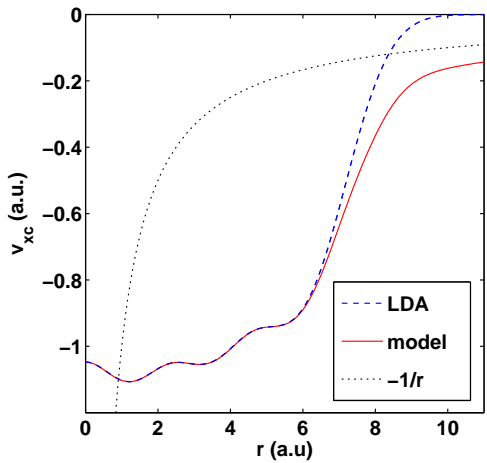}
\caption{(Color online) Exchange-correlation potentials
of a 42-electron parabolic quantum dot calculated
with the LDA (dashed line) and the present model (solid line).
The dotted line shows the desired asymptotic behavior ($-1/r$).
}
\label{fig2}
\end{figure}

Figure~\ref{fig2}
shows the xc potentials for a significantly larger parabolic 
QD with 42 electrons. The confinement strength
is now $\omega=0.5$ (Ref.~\onlinecite{units}).
Here we have performed a self-consistent
LDA calculation using the {\tt octopus} code~\cite{octopus}
and use the LDA density as an input for the model potential.
Again the correct asymptotic limit seems to be recovered
by the model potential,
although the tail cannot be followed
as far as in the exact two-electron case in Fig.~\ref{fig1}.
This is due to the numerical instability in the implementation
of the LDA correlation in the regime where 
the density is very small
($\lesssim 10^{-16}$). Nevertheless, Fig.~\ref{fig2}
confirms the improvement of the model potential 
over the LDA for a large QD system.

\begin{figure}
\includegraphics[width=0.75\columnwidth]{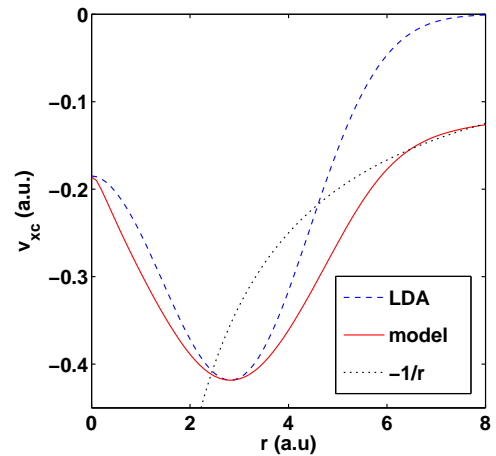}
\caption{(Color online) Exchange-correlation potentials
of a two-electron parabolic ($\omega = 0.5$, $r_0=3$) quantum ring calculated
with the LDA (dashed line) and the present model (solid line).
The dotted line shows the desired asymptotic behavior ($-1/r$).
}
\label{fig3}
\end{figure}

\begin{figure}
\includegraphics[width=0.75\columnwidth]{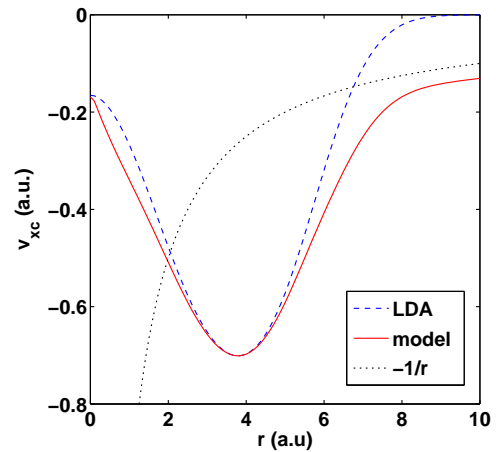}
\caption{(Color online) As in Fig.~\ref{fig3} but for ten electrons.}
\label{fig4}
\end{figure}

Finally, we consider quantum rings 
defined by an external potential 
$v_{\rm ext}(r)=\omega^2 (r-r_0)^2/2$, where
$\omega=0.5$ and the ring radius $r_0=3$. 
The obtained potentials for two-electron
and ten-electron rings are shown in Figs.~\ref{fig3} and \ref{fig4},
respectively.
Similarly to the previous test cases, the correct asymptotic 
behavior ($-1/r$) is obtained with the model potential
for both quantum rings. At the ring radius 
where the density gradient is zero, the model
agrees with the LDA in a smooth fashion.

It is noteworthy that the present model for the xc potential 
sets specific constraints on the external potential. In particular,
infinite potentials (at finite positions)
-- leading to situations where {\em both} the
density and its gradient are zero -- may result in divergence
of $F(x)$. A typical example would be a quantum ring studied
in Ref.~\onlinecite{ring}. In this respect, the present potential
needs to be applied with special care, or alternative
models should be considered, when studying 2D structures
represented by such external potentials. Nevertheless, 
the presented scheme opens up a path to
design of a more general GGA approach.

\section{Summary}

In summary, we have derived a generalized-gradient 
approximation for the exchange-correlation potential 
with a proper asymptotic decay for two-dimensional nanostructures.
The expression has a proper asymptotic decay
and thus corrects one of the most significant errors in the
commonly used two-dimensional local-density approximation.
The validity of the model potential is confirmed in 
quantum dots and rings with different number of electrons. 
We believe that the presented approach gives promising prospects
in two-dimensional many-body physics,
first and foremost for systems where the correct decay
of the exchange-correlation potential is of great importance, for example,
quantum-dot molecules and chains, coupled quantum rings, and
periodic electron lattices such as the artificial
graphene.~\cite{2DEGgraphene}

\begin{acknowledgments}
We thank Robert van Leeuwen for useful discussions.
This work has been  supported by DOE grant DE-FG02-05ER46203 (S.P.)
and by the Academy of Finland (E.R.).
\end{acknowledgments}


\begin{thebibliography}{ll}
\bibitem{note1}
We emphasize that the very same behavior is valid for atoms and molecules.
\bibitem{note2}
In our previous works (see, e.g., Refs.~\onlinecite{x1,ring,gamma,c1,c2,lccs}), we have verified 
that functionals derived from models of wave functions, or of related
quantities with a Gaussian-like behavior, account well for 
a variety of semiconductor heterostructures (quantum dots, rings, and slabs).
\end{thebibliography}
\end{document}